\documentclass{article}
\def\~{\tilde}
\usepackage{graphicx}

\begin{document}
\title{\bf{Experiment can decide between\\ standard and Bohmian quantum mechanics}}
\author{M. Golshani \footnote{E-mail: golshani@ihcs.ac.ir} \hspace{1mm} and O. Akhavan \footnote{E-mail: akhavan@mehr.sharif.ac.ir
}}
\date{\small{{\it Department of Physics, Sharif University of Technology,
P.O. Box 11365--9161, Tehran,  Iran\\Institute for Studies in
Theoretical Physics and Mathematics, P.O. Box 19395--5531,
Tehran, Iran}}} \maketitle

\begin{abstract}
In this investigation, we have considered two thought experiments
to make a comparison between predictions of the standard and the
Bohmian quantum mechanics.  Concerning this, a two-particle
system has been studied at two various situations of the entangled
and the unentangled states. In the first experiment, the two
theories can predict different results at the individual level,
while their statistical results are the same. In the other
experiment, not only they are in disagreement at the individual
level, but their equivalence at the statistical level also breaks
down, if one uses selective detection.  Furthermore, we discuss
about some objections that can be raised against the results of
the two suggested experiments.
\end{abstract}


PACS number: 03.65.Bz\\
{\it Keywords}: Bohmian quantum mechanics, Thought experiment,
Entanglement, Selective detection, Inconsistent prediction

\section{Introduction}

The statistical interpretation of the wave function of the
standard quantum mechanics (SQM) is consistent with all performed
experiments.  An interference pattern on a screen is built up by
a series of apparently random events, and the wave function
correctly predicts where the particle is most likely to land in an
ensemble of trials.  Instead, one may take the view that the
characteristic distribution of spots on a screen which build up
an interference pattern is an evidence that the wave function has
a more potent physical role.  If one attempts to understand the
experimental results as the outcome of a causally connected
series of individual process, then one is free to inquire about
further significance of the wave function and to introduce other
concepts in addition to the wave function.  Bohm {\cite{Boh}}, in
1952, showed that an individual physical system comprises a wave
propagating in space-time together with a point particle which
moves continuously under the guidence of the wave [1-4]. He
applied his theory to a range of examples drawn from
non-relativistic quantum mechanics and speculated on the possible
alternations in the particle and field laws of motion such that
the predictions of the modified theory continue to agree with
those of SQM where this is tested, but it could disagree in as yet
unexplored domains {\cite{Hol}}. For instance, when Bohm presented
his theory in 1952, experiments could be done with an almost
continuous beam of particles.  Thus, it was impossible to
discriminate between the standard and the Bohmian quantum
mechanics (BQM) at the individual level.  However, the two
theories can be discriminated at this level, because SQM is a
probabilistic theory while BQM is a precisely defined and
deterministic theory.

  In recent years, the significance of proposals that can predict different
results for SQM and BQM have been the subject of many discussions
[for example, 5-17]. At first, it seems that definition of time
spent by a particle within a classically forbidden barrier
provides a good evidence for the preference of BQM.  But, there
are difficult technical questions, both theoretically and
experimentally, that are still unsolved about these tunneling
times [4,5].  On the other hand, Englert et al. {\cite{Eng}} and
Scully {\cite{Scu}} have claimed that in some cases Bohm's
approach gives results that disagree with those obtained from SQM
and, in consequence, with experiment. Concerning this, at first
Dewdney et al. {\cite{Dew}} and then Hiley et al. {\cite{Hil}}
showed that the specific objections raised by Englert and Scully
cannot be sustained. Furthermore, Hiley et al. {\cite{Hil}}
believe that no experiment can decide between the standard
interpretation and Bohm's interpretation. However, Vigier
{\cite{Vig}}, in his recent work, has given a brief list of new
experiments which suggest that the U(1) invariant massless photon
assumed properties of light within the standard interpretation,
are too restrictive and that the O(3) invariant massive photon
causal de Broglie-Bohm interpretation of quantum mechanics, is now
supported by experiments.  In addition, Leggett {\cite{Leg}}
considered some thought experiments involving macrosystems which
can predict different results for SQM and BQM. \footnote{Legget
{\cite{Leg}} assumes that the experimental predictions of SQM will
continue to be realized under the extreme conditions specified,
although a test of this hypothesis is part of the aim of the
macroscopic quantum cohrence program.  In addition, he considered
BQM as another interpretation of the same theory rather than an
alternative theory.} Furthermore, in some of the recent works,
feasible thought experiments have been suggested to distinguish
between SQM and BQM [12,13,17].  In one of the works, Ghose
{\cite{Gho}} indicated that although BQM is equivalent to SQM
when the averages of dynamical variables are taken over a Gibbs
ensemble of Bohmian trajectories, the equivalence breaks down for
ensembles built over clearly separated short intervals of time in
special entangled two-bosonic particle systems.  In another work
{\cite{Gol}}, we have shown incompatibility between SQM and BQM at
the individual and ensemble levels, using a two-slit device whose
source emits two uncorrelated identical particles.  However,
Marchildon [14,15] has tried to show that there is no reason to
expect discrepancies between BQM and SQM in the context of the
two-particle interference devices.  Ghose {\cite{Ghos}} believes
that Marchildon's arguments against his work are untenable and
that his basic conclusion stands.  In addition, we have shown
elsewhere {\cite{Gols}} that the incompatibility between SQM and
BQM can also appeare in a new more feasible experiment at the
individual level and that the statistical disagreement is also
valid if we use our selective detection. It should be noted that,
the role of selective detection in altering the statistics of
predictions is explained by Durr et al. {\cite{Dur}}.

In this work, in parallel to the works [12,13], we have studied
the entangled and the unentangled wave functions that can be
imputed to a two-particle interference device, using a Gaussian
wave function as a real representation.  Then, SQM and BQM
predictions are compared at both the individual and the
statistical level. \footnote{The individual level refers to our
experiment with pairs of particles which are emitted in clearly
separated short intervals of time, and by statistical level we
mean our final interference pattern.}  We also discuss about some
objections that can be raised.

\section{Description of two-particle double-slit experiment}

Consider the famous double-slit experiment.  Instead of the usual
one-particle emitting source, one can consider a special point
source $S_{1}$, so that a pair of identical non-relativistic
particles originate simultaneously from it.  We assume that, the
intensity of the beam is so low that at a time we have only a
single pair of particles passing through the slits and the
detection screen $S_{2}$ registers only those pairs of particles
that reach it simultaneously, so that the interference effects of
single particles will be eliminated. Furthermore, it is assumed
that the detection process has no causal role in the phenomenon
of interference {\cite{Hol}}.  In the two-dimensional coordinate
system, the centers of the two slits are located at $(0,\pm Y)$.

Concerning the assumed source, we can have two alternatives:\\
{\it 1.} The wave function of the two emitted particles are so
entangled that if one particle passes from the upper (lower)
slit, the other particle must go through lower (upper) slit.  In
 other words, the total momentum and the center of mass of the
two particles in the
$y$-direction is zero at the source, that is, $p_{1y}-p_{2y}=0$ and $y_1+y_2=0$.\\
{\it 2.} The wave function of the two emitted particles have no
correlation and they are unentangled.  In other words, the
emission of each particle is done freely and two particles can be
considered independently.\\
In the following, we shall study each one of the two alternatives,
separately, and apply them, using SQM.  Then, in the next section,
the Bohmian predictions have been compared with those of SQM.

\subsection{Entangled wave function}

We take the wave incident on the double-slit screen to be a plane
wave of the form
\begin{equation}
\psi_{in}(x_{1},y_{1};x_{2},y_{2};t)=ae^{i[k_{x}(x_{1}+x_{2})+k_{y}(y_{1}-y_{2})]}e^{-iEt/\hbar},
\end{equation}
where $a$ is a constant and
$E=E_{1}+E_{2}=\hbar^{2}(k_{x}^{2}+k_{y}^{2})/m$ is the total
energy of the system of the two identical particles.  The
parameter $m$ is the mass of each particle and $k_{i}$ is the
wave number of particle in the $i$-direction.  For mathematical
simplicity, we avoid slits with sharp edges which produce the
mathematical complexity of Fresnel diffraction, i.e., we assume
that the slits have soft edges, so that the Gaussian wave packets
are produced along the $y$-direction, and that the plane wave
along the $x$-axis remain unchanged {\cite{Hol}}.  We take the
time of the formation of the Gaussian wave to be $t=0$. Then, the
emerging wave packets from the slits $A$ and $B$ are respectively
\begin{equation}
\psi_{A}(x,y)=a(2\pi\sigma_{0}^{2})^{-1/4}e^{-(y-Y)^{2}/4\sigma_{0}^{2}}e^{i[k_{x}x+k_{y}(y-Y)]},
\end{equation}
\begin{equation}
\psi_{B}(x,y)=a(2\pi\sigma_{0}^{2})^{-1/4}e^{-(y+Y)^{2}/4\sigma_{0}^{2}}e^{i[k_{x}x-k_{y}(y+Y)]},
\end{equation}
where $\sigma_{0}$ is the half-width of each slit.

 Now, for this
two-particle system, the total wave function at the detection
screen $S_{2}$, at time $t$, can be written as
\begin{eqnarray}
\psi(x_{1},y_{1};x_{2},y_{2};t) =
N[\psi_{A}(x_{1},y_{1},t)\psi_{B}(x_{2},y_{2},t)\pm\psi_{A}(x_{2},y_{2},t)\psi_{B}(x_{1},y_{1},t)],
\end{eqnarray}
where $N$ is a normalization constant which is unimportant here,
and
\begin{equation}
\psi_{A}(x,y,t)=a(2\pi
\sigma_{t}^{2})^{-1/4}e^{-(y-Y-u_{y}t)^{2}/4\sigma_{0}\sigma_{t}}e^{i[k_{x}x+k_{y}(y-Y-u_{y}t/2)-E_{x}t/\hbar]},
\end{equation}
\begin{equation}
\psi_{B}(x,y,t)=a(2\pi
\sigma_{t}^{2})^{-1/4}e^{-(y+Y+u_{y}t)^{2}/4\sigma_{0}\sigma_{t}}e^{i[k_{x}x-k_{y}(y+Y+u_{y}t/2)-E_{x}t/\hbar]},
\end{equation}
where
\begin{equation}
\sigma_{t}=\sigma_{0}(1+\frac{i\hbar t}{2m\sigma_{0}^{2}}),
\end{equation}
and
\begin{eqnarray}
&& u_{y}=\frac{\hbar k_{y}}{m},\cr&&
E_{x}=\frac{1}{2}mu_{x}^{2}.
\end{eqnarray}
Note that, the upper and lower signs in the total entangled wave
function (4) are due to symmetric and anti-symmetric wave
function under the exchange of particles 1 and 2, corresponding
to bosonic and fermionic property, respectively.

\subsection{Unentangled wave function}

In this case, the incident plane wave can be considered to be
\begin{equation}
\tilde{\psi}_{in}(x_{1},y_{1};x_{2},y_{2};t)=ae^{i[k_{x}(x_{1}+x_{2})+k_{y}(\pm
y_{1}\pm y_{2})]}e^{-iEt/\hbar},
\end{equation}
where it has four cases corresponding to each sign.  Now, for this
two-particle system, the total wave function at time $t$ can be
written as
\begin{eqnarray}
\lefteqn{\tilde{\psi}(x_{1},y_{1};x_{2},y_{2};t) =}\cr&&
\widetilde{N}[\psi_{A}(x_{1},y_{1},t)\psi_{B}(x_{2},y_{2},t)+\psi_{A}(x_{2},y_{2},t)\psi_{B}(x_{1},y_{1},t)\cr&&
 +\psi_{A} (x_{1},y_{1},t)\psi_{A}(x_{2},y_{2},t)+\psi_{B}
(x_{1},y_{1},t)\psi_{B} (x_{2},y_{2},t)]\cr&&
=\widetilde{N}[\psi_{A}(x_{1},y_{1},t)+\psi_{B}(x_{1},y_{1},t)][\psi_{A}(x_{2},y_{2},t)+\psi_{B}(x_{2},y_{2},t)],
\end{eqnarray}
where $\widetilde{N}$ is another normalization constant.

\subsection{SQM's prediction}

Based on SQM, the wave function can be associated with an
individual physical system.  It provides the most complete
description of the system that is, in principle, possible.  The
nature of description is statistical, and concerns the
probabilities of the outcomes of all conceivable measurements
that may be performed on the system.  It is well known from SQM
that, the probability of simultaneous detection of the particles
at $y_{M}$ and $y_{N}$, at the screen $S_{2}$, located at
$x_{1}=x_{2}=D$ and $t=D/u_{x}$, is equal to
\begin{equation}
P_{12}(y_{M},y_{N},t)=\int_{y_{M}}^{y_{M}+\triangle}dy_{1}\int_{y_{N}}^{y_{N}+\triangle}dy_{2}|\psi(x_{1},y_{1};x_{2},y_{2};t)|^{2}.
\end{equation}
The parameter $\Delta$, which is taken to be small,  is a measure
of the size of the detectors.  We shall compare this prediction
of SQM with that of BQM.

\section{Bohmian predictions and their comparison with SQM}

Based on basic postulates of BQM, an individual physical system
consists of a wave propagating in space-time and a point particle
which moves continuously under the guidance of the wave. The wave
function $\psi(\overrightarrow{x},t)$ is a solution of
Schr$\ddot{o}$dinger's equation and the particle motion is
obtained from the following first order differential equation
\begin{equation}
\overrightarrow{\dot{x}_{i}}(\overrightarrow{x},t)=\frac{1}{m_{i}}
\overrightarrow{\nabla_{i}}S(\overrightarrow{x},t)=\frac{\hbar}{m_{i}}Im\left(
\frac{\overrightarrow{\nabla_{i}}\psi(\overrightarrow{x},t)}{\psi(\overrightarrow{x},t)}\right),
\end{equation}
where $\overrightarrow{x}=(\overrightarrow{x_1}
,\overrightarrow{x_2},..., \overrightarrow{x_n})$, and
$S(\overrightarrow{x},t)$ is the phase of
$\psi(\overrightarrow{x},t)$ in polar form, that is,
\begin{equation}
\psi(\overrightarrow{x},t)= R(\overrightarrow{x},t)
e^{iS(\overrightarrow{x},t)/\hbar}.
\end{equation}
To compare between the two theories, here, we study the speed of
particles 1 and 2 in the $y$-direction, that is,
\begin{equation}
\dot{y}_{1}(x_{1},y_{1};x_{2},y_{2};t)=\frac{\hbar}{m}Im(\frac{\partial_{y_{1}}\psi(x_{1},y_{1};x_{2},y_{2};t)
}{\psi(x_{1},y_{1};x_{2},y_{2};t)}),
\end{equation}
\begin{equation}
\dot{y}_{2}(x_{1},y_{1};x_{2},y_{2};t)=\frac{\hbar}{m}Im(\frac{\partial_{y_{2}}\psi(x_{1},y_{1};x_{2},y_{2};t)
}{\psi(x_{1},y_{1};x_{2},y_{2};t)}).
\end{equation}
Remember that two kinds of the wave function could be considered;
the entangled and the unentangled wave function.  Thus, in the
following we study each of them, separately.

\subsection{Predictions for the entangled wave function}

Consider the entangled wave function (4).  By substituting it in
(14) and (15), we have
\begin{eqnarray}
\dot{y}_{1}=&N\frac{\hbar}{m}&Im\{
\frac{1}{\psi}[[-2(y_{1}-Y-u_{y}t)/4\sigma_{0}\sigma_{t}+ik_{y}]\psi_{A_{1}}\psi_{B_{2}}\cr
           &\pm&[-2(y_{1}+Y+u_{y}t)/4\sigma_{0}\sigma_{t}-ik_{y}]\psi_{A_{2}}\psi_{B_{1}}]\},
\end{eqnarray}
\begin{eqnarray}
\dot{y}_{2}=&N\frac{\hbar}{m}&Im\{\frac{1}{\psi}[[-2(y_{2}+Y+u_{y}t)/4\sigma_{0}\sigma_{t}-ik_{y}]\psi_{A_{1}}\psi_{B_{2}}\cr
           &\pm&[-2(y_{2}-Y-u_{y}t)/4\sigma_{0}\sigma_{t}+ik_{y}]\psi_{A_{2}}\psi_{B_{1}}]\},
\end{eqnarray}
On the other hand, from (5) and (6) one can see that,
\begin{eqnarray}
&&\psi_{A}(x_{1},y_{1},t)=\psi_{B}(x_{1},-y_{1},t),\cr
&&\psi_{A}(x_{2},y_{2},t)=\psi_{B}(x_{2},-y_{2},t),
\end{eqnarray}
 which indicate the reflection symmetry  of $\psi(x_{1},y_{1};x_{2},y_{2};t)$ with
respect to the $x$--axis. Using this symmetry in (16) and (17), we
have
\begin{eqnarray}
&&\dot{y}_{1}(x_{1},y_{1};x_{2},y_{2};t)=\mp\dot{y}_{1}(x_{1},-y_{1};x_{2},-y_{2};t),\cr&&
\dot{y}_{2}(x_{1},y_{1};x_{2},y_{2};t)=\mp\dot{y}_{2}(x_{1},-y_{1};x_{2},-y_{2};t).
\end{eqnarray}
These relations show that if $y_{1}(t)=y_{2}(t)=0$, i.e., two
particles are on the $x$-axis, simultaneously, then the speed of
each bosonic particles in the $y$-direction is zero along the
symmetry axis $x$, but we have no such  constraint on fermionic
particles, as was mentioned by Ghose {\cite{Gho}}.  We have shown
elsewhere {\cite{Gols}} that, there is such a constraint on both
bosonic and fermionic particles, using the two entangled
particles in a two double-slit device.

If we consider $y=(y_{1}+y_{2})/2$ to be the vertical coordinate
of the center of mass of the two particles, then we can write
\begin{eqnarray}
\dot{y}&=&(\dot{y}_{1}+\dot{y}_{2})/2\cr
       &=&N\frac{\hbar}{2m}Im\{\frac{1}{\psi}(-\frac{y_{1}+y_{2}}{2\sigma_{0}\sigma_{t}})(\psi_{A_{1}}\psi_{B_{2}}\pm\psi_{A_{2}}\psi_{B_{1}})\}\cr
       &=&\frac{(\hbar/2m\sigma_{0}^{2})^{2}}{1+(\hbar/2m\sigma_{0}^{2})^{2}t^{2}}yt.
\end{eqnarray}
Solving this differential equation, we get the path of the
$y$-coordinate of the center of mass
\begin{equation}
y=y_{0}\sqrt{1+(\hbar/2m\sigma_{0}^{2})^{2}t^{2}}.
\end{equation}
Using equation (21) and doing the same as what was done in ref.
[17], one obtains the quantum potential for the center of mass
motion
\begin{equation}
Q_{cm}=\frac{my_{0}^{4}}{2y^{2}}(\frac{\hbar}{2m\sigma_{0}^{2}})^{2}=\frac{1}{2}my_{0}^{2}\frac{(\hbar/2m\sigma_{0}^{2})^{2}}{1+(\hbar
t/2m\sigma_{0}^{2})^{2}}.
\end{equation}
If the center of mass of the system is exactly on the $x$-axis at
$t=0$ , then $y_{0}=0$, and the center of mass of the system will
always remain on the $x$-axis.  In addition, the quantum
potential for the center of mass of the two particles is zero at
all times. Thus, we have $y_{1}(t)=-y_{2}(t)$ and the two
particles, in both the bosonic and fermionic case,  will be
detected at points symmetric with respect to the $x$-axis.  This
differs from the prediction of SQM, as the probability relation
(11) shows. SQM predicts that the probability of asymmetrical
detection of the pair of particles can be different from zero in
contrast to BQM's symmetrical prediction. Furthermore, according
to SQM's prediction, the probability of finding two particles at
one side of the $x$-axis can be non-zero while it is shown that
BQM forbids such events, provided that $y_{0}=0$. Figure 1 shows
one of the typical inconsistencies which can be considered at the
individual level. Based on BQM, bosonic and fermionic particles
have the same results, but, we know that if one bosonic particle
passes through the upper (lower) slit, it must detected on the
upper (lower) side on the $S_{2}$ screen, due to relations (19).
Instead, there is no such  constraint on fermionic particles.
\begin{center}
\begin{figure}[t]
\includegraphics[width=15cm,height=7cm,angle=0]{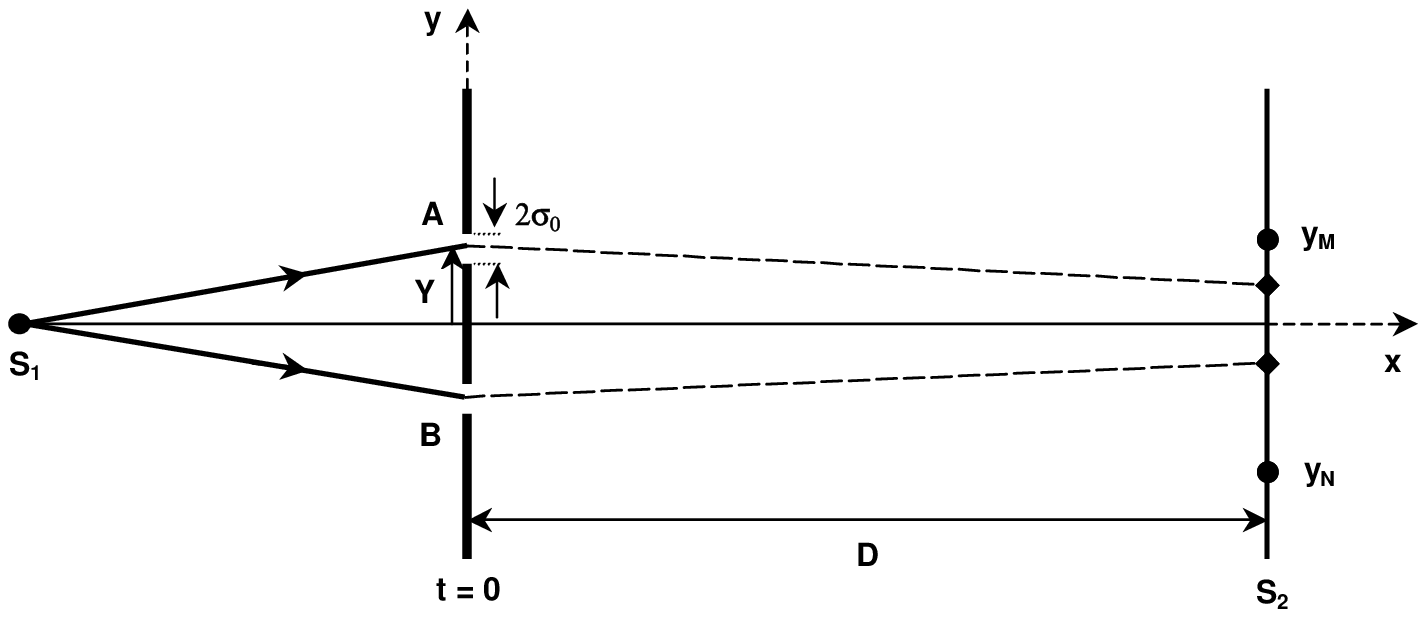}
\caption{A two-slit experiment in which two identical entangled
particles are emitted from the source $S_{1}$.  Then, they pass
through the slits $A$ and $B$, and finally they are detected on
the screen $S_{2}$, simultaneously. We assumed that, $y_{0}=0$,
or $\langle y_{0}\rangle=0$ under $\triangle y_{0}\ll\sigma_{0}$
and $\hbar t/2m\sigma_{0}^{2}\sim 1$ conditions. It is clear that
dashed lines are not real trajectories.}
\end{figure}
\end{center}

  We assumed that the two particles
are entangled so that in spite of a position distribution for
each particle, $y_{0}$ can be always considered to be on the
$x$-axis. However, one may argue that, it is necessary to consider
a position distribution for $y_{0}$, that is, $\triangle
y_{0}\neq0$ while $\langle y_{0}\rangle=0$. Therefore, it may
seem that, not only symmetrical detection of the two particles is
violated, but also they can be found at one side of the $x$-axis
on the $S_{2}$ screen, because the majority of pairs can not be
simultaneously on the $x$-axis {\cite{Marc}}.  Ghose
{\cite{Ghos}} believes that the two entangled bosonic particles
cannot cross the symmetry axis even if  we have the situation
$(y_1+y_2)_{t=0}\neq 0$. However, by accepting Marchildon's
argument about this situation {\cite{Marc}}, this problem can be
solved if we adjust $\triangle y_{0}$ to be very small and $\hbar
t/2m\sigma_{0}^{2}\sim 1$.  We assume that, to maintain
symmetrical detection about the $x$-axis with reasonable
approximation, the center of mass deviation from the $x$-axis
must be smaller than the distance between any two neighbouring
maxima on the screen $S_{2}$, that is,
\begin{eqnarray}
\triangle y\ll\frac{\lambda D}{2Y}\simeq\frac{\pi\hbar t}{Ym},
\end{eqnarray}
where $\lambda$ is the de Broglie wavelength.  For conditions
$\hbar t/2m\sigma_{0}^{2}\sim 1$, $Y\sim\sigma_{0}$ and using
equation (21), one obtains
\begin{eqnarray}
\triangle y_{0}\ll\frac{\pi\hbar t}{Ym}\sim\sigma_{0}.
\end{eqnarray}
Therefore, if we use a source with $\triangle y_{0}\ll\sigma_{0}$,
we will obtain $y\sim y_{0}\ll\sigma_{0}$ for each individual
observation, and our symmetrical detection can be maintained with
good approximation.  In this case, we only lose our information
about the trajectory of bosonic particles.  It is evident that, if
one considers $\triangle y_{0}\sim\sigma_{0}$, as was done in
[15], the incompatibility between the two theories will
disappear. But based on the entanglement of the two particles in
the $y$-direction, we believe that, instead of the usual
one-particle double-slit experiment with $\triangle
y_{0}\sim\sigma_{0}$, our two-particle system provides a new
situation in which we can adjust $y_{0}$ independent of
$\sigma_{0}$, so that
\begin{eqnarray}
0\leq y_{0}=\frac{1}{2}(y_{1}+y_{2})_{t=0}\ll\sigma_{0}.
\end{eqnarray}
Although it is obvious that $(\triangle y_{1})_{t=0}=(\triangle
y_{2})_{t=0}\sim\sigma_{0}$, but the position entanglement of the
two particles at the source $S_{1}$ in the $y$-direction makes
them always satisfy equation (25), which is not feasible in the
one-particle double-slit devices.

Now, one can compare the results of SQM and BQM at the ensemble
level. To do this, we consider an ensemble of pairs of particles
that have arrived at the detection screen $S_{2}$ at different
times $t_{i}$.  It is well known that, in order to ensure the
compatibility between SQM and BQM for ensemble of particles, Bohm
added a further postulate to his three basic and consistent
postulates [1-3].  Based on this further postulate, the
probability that a particle in the ensemble lies between
$\overrightarrow{x}$ and $\overrightarrow{x}+d\overrightarrow{x}$,
at time $t$, is given by
\begin{eqnarray}
P(\overrightarrow{x},t)=R^{2}(\overrightarrow{x},t).
\end{eqnarray}
Thus, the joint probability of simultaneous detection for all
pairs of particles of the ensemble arriving at $S_{2}$ is
\begin{eqnarray}
P_{12}=\lim_{N\rightarrow\infty}\sum_{i=1}^{N}R^{2}(y_{1}(t_{i}),-y_{1}(t_{i}),t_{i})\equiv\int^{+\infty}_{-\infty}dy_{1}\int^{+\infty}_{-\infty}dy_{2}|\psi(y_{1},y_{2},t)|^{2}=1,
\end{eqnarray}
where every term in the sum shows only one pair arriving on the
screen $S_{2}$ at the symmetrical points about the $x$-axis at
time $t_{i}$, with the intensity of $R^{2}$.  If all times $t_{i}$
in the sum are taken to be $t$, the summation on $i$ can be
converted to an integral over all paths that cross the screen
$S_{2}$ at that time, and we obtain an interference pattern.
Then, one can consider the joint probability of detecting two
particles at two arbitrary points $y_{M}$ and $y_{N}$ which can
belong to different pairs
\begin{equation}
P_{12}(y_{M},y_{N},t)=\int_{y_{M}}^{y_{M}+\Delta}dy_{1}\int_{y_{N}}^{y_{N}+\Delta}dy_{2}|\psi(y_{1},y_{2},t)|^{2},
\end{equation}
which is similar to the prediction of SQM, but obtained in a
Bohmian way, as was shown by Ghose {\cite{Gho}}.  Thus, it appears
that for such conditions, the possibility of distinguishing the
two theories at the statistical level is impossible, as was
expected [1-3, 9, 18].

Here, to show equivalence of the two theories, we have assumed
for simplicity that $y_{0}=0$.  If one consider $y_{0}\neq 0$ or
$\triangle y_{0}\neq 0$, the equivalence of the two theories is
maintained, as it is argued by Marchildon {\cite{Marc}}. But,
using this special case, we show that assumption of $y_{0}=0$ is
consistent with statistical results of SQM and in consequence,
finding such a source may not be impossible.

\subsection{Predictions for the unentangled wave function}

In this subsection, we complete our discussion by considering the
unentangled wave function (10) and some of the Marchildon's hints
{\cite{Marc}}. Based on equations (14) and (15), Bohmian
velocities of particle 1 and 2 can be obtained as
\begin{eqnarray}
\dot{y}_{1}=&\widetilde{N}&\frac{\hbar}{m}Im\{\frac{1}{(\psi_{A_{1}}+\psi_{B_{1}})}([\frac{-2(y_{1}-Y-u_{y}t)}{4\sigma_{0}\sigma_{t}}+ik_{y}]\psi_{A_{1}}\cr
           &+&[\frac{-2(y_{1}+Y+u_{y}t)}{4\sigma_{0}\sigma_{t}}-ik_{y}]\psi_{B_{1}}\},
\end{eqnarray}
\begin{eqnarray}
\dot{y}_{2}=&\widetilde{N}&\frac{\hbar}{m}Im\{\frac{1}{(\psi_{A_{2}}+\psi_{B_{2}})}([\frac{-2(y_{2}-Y-u_{y}t)}{4\sigma_{0}\sigma_{t}}+ik_{y}]\psi_{A_{2}}\cr
           &+&[\frac{-2(y_{2}+Y+u_{y}t)}{4\sigma_{0}\sigma_{t}}-ik_{y}]\psi_{B_{2}}\},
\end{eqnarray}
Thus, as we expected, the speed of each particle is independent
of the other.  Using these relations as well as equations (18), we
have
\begin{eqnarray}
&&\dot{y}_{1}(x_{1},y_{1},t)=-\dot{y}_{1}(x_{1},-y_{1},t),\cr&&
\dot{y}_{2}(x_{2},y_{2},t)=-\dot{y}_{2}(x_{2},-y_{2},t).
\end{eqnarray}
This implies that the $y$-component of the velocity of each
particle would vanish on the $x$-axis.  Although these relations
are similar to the relations that were obtained for the entangled
wave function, but here we have an advantage: none of the
particles can cross the $x$-axis nor are tangent to it,
independent of the other particle's position.  This property can
be used to show that BQM's predictions are incompatible with
SQM's.

  To see this
incompatibility, we use a special detection process on the screen
$S_{2}$ that we call it selective detection.  In the selective
detection, we register only those pair of particles which are
detected on the two sides of the $x$-axis, simultaneously.  That
is, we eliminate the cases of detecting only one particle or
detecting both particles of the pair on the upper or lower part
of the $x$-axis on the screen.  Again, it is useful to obtain the
equation of motion of the center of mass in the $y$-direction.
Using equation (29) and (30), one can show that,
\begin{eqnarray}
\dot{y}=\frac{(\hbar/2m\sigma_{0}^{2})^{2}t(y_{1}+y_{2})/2}{1+(\hbar/2m\sigma_{0}^{2})^{2}t^{2}}
+\widetilde{N}\frac{\hbar}{2m}Im\{\frac{1}{\psi}(\frac{Y+u_{y}t}{\sigma_{0}\sigma_{t}}+2ik_{y})(\psi_{A_{1}}\psi_{A_{2}}-\psi_{B_{1}}\psi_{B_{2}})\}.
\end{eqnarray}
We can assume that the distance between the source and the
two-slit screen is so large that we have $k_y\simeq 0$.  Then,
using the special case $Y\ll\sigma_0$, the second term in eq. (32)
becomes negligible and the equation of motion for the
$y$-coordinate of the center of mass is reduced to
\begin{equation}
\dot{y}\simeq\frac{(\hbar/2m\sigma_{0}^{2})^{2}}{1+(\hbar/2m\sigma_{0}^{2})^{2}t^{2}}yt,
\end{equation}
and similar to the last experiment, we have
\begin{equation}
y\simeq y_{0}\sqrt{1+(\hbar/2m\sigma_{0}^{2})^{2}t^{2}}.
\end{equation}
Since for this special source there was not any entanglement
between the two particles, we must have $\triangle
y_0\sim\sigma_0$. Consider the case in which $\langle
y_0\rangle=0$.  To obtain symmetrical detection with reasonable
approximation, we consider $\hbar t/2m\sigma_{0}^{2}\sim 1$ so
that $y\sim y_0$.  Then, according to eq. (23), one can write
\begin{equation}
\triangle y_0\ll\frac{\pi\hbar t}{Ym},
\end{equation}
which yields
\begin{equation}
Y\ll 2\pi\sigma_0.
\end{equation}
Therefore, under these conditions, BQM's symmetrical prediction is
incompatible with SQM's asymmetrical prediction.  Figure 2 shows
a schematic drawing of BQM's symmetrical detection occurred at
the first maximum for the condition $Y\ll 2\pi\sigma_0$.
\begin{center}
\begin{figure}[t]
\includegraphics[width=15cm,height=10cm,angle=0]{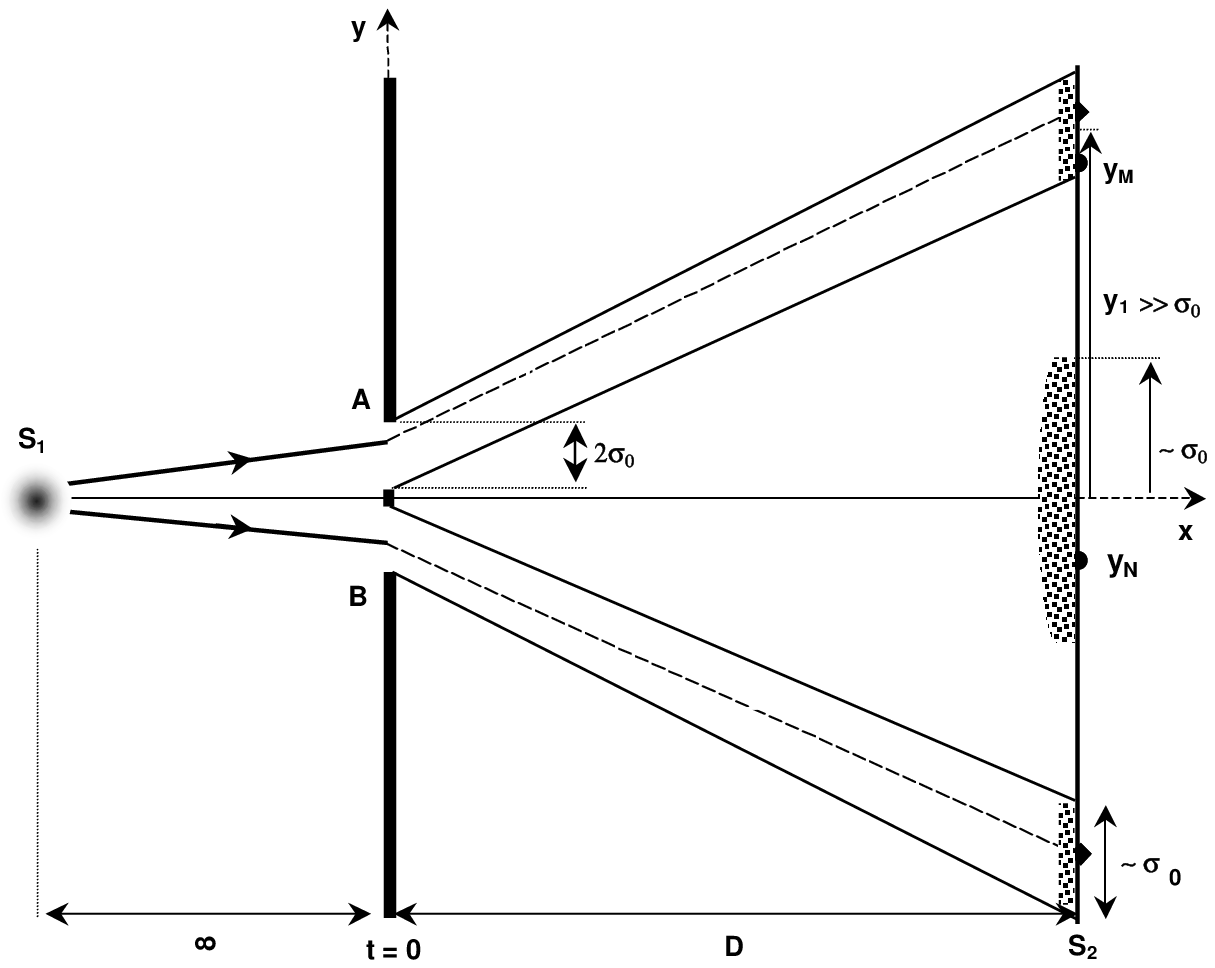}
\caption{Schematic drawing of a two-slit experiment in which two identical
unentangled particles are emitted by the source $S_1$.  The
symmetrical detection is not predicted at the central maxima,
using both SQM and BQM.  But, we have symmetrical detection at
the other maxima (for example, at $y_1$ as the first maxima)
under the condition $Y\ll 2\pi\sigma_0$, using BQM.}
\end{figure}
\end{center}

  Now, consider conditions under which $y_{0}\neq 0$
and $\hbar t/2m\sigma_{0}^{2}\gg 1$.  Then, the $x$-axis will not
be an axis of symmetry and we have a new point on the $S_{2}$
screen along the $y$-axis around which all pairs of particles will
be detected symmetrically.  Thus, based on BQM, that is relations
(31) and (34), there will be an empty interval
\begin{equation}
L=2y\simeq \frac{\hbar ty_{0}}{m\sigma_{0}^{2}},
\end{equation}
on the screen $S_{2}$ where no particle is recorded. But, one can
argue that the quantum distribution of the two particles does not
allow to form the empty interval on the screen. Thus, assume that
$\langle y_0\rangle\neq 0$ and $\triangle y_0\sim\sigma_0$. Then,
we can have a relative empty interval of low intensity particles
that has a length
\begin{equation}
L\simeq 2\langle y\rangle\simeq\frac{\hbar
t}{m\sigma_{0}^{2}}\langle y_{0}\rangle,
\end{equation}
if $\triangle y\ll L$ condition is satisfied.  It is obvious
that, the last condition corresponds to $\triangle y_0\ll \langle
y_{0}\rangle$.  Therefore, it is seen that, using BQM and under
the condition
\begin{equation}
Y\ll\sigma_0\ll\langle y_0\rangle,
\end{equation}
 a considerable change in the position of the source $S_{1}$ toward
positive (negative) $y$-direction produces a region with very low
intensity in the interference pattern above (below) the $x$-axis
which cannot be predicted by SQM, as is shown in Figure 3.
\begin{center}
\begin{figure}[t]
\includegraphics[width=15cm,height=10cm,angle=0]{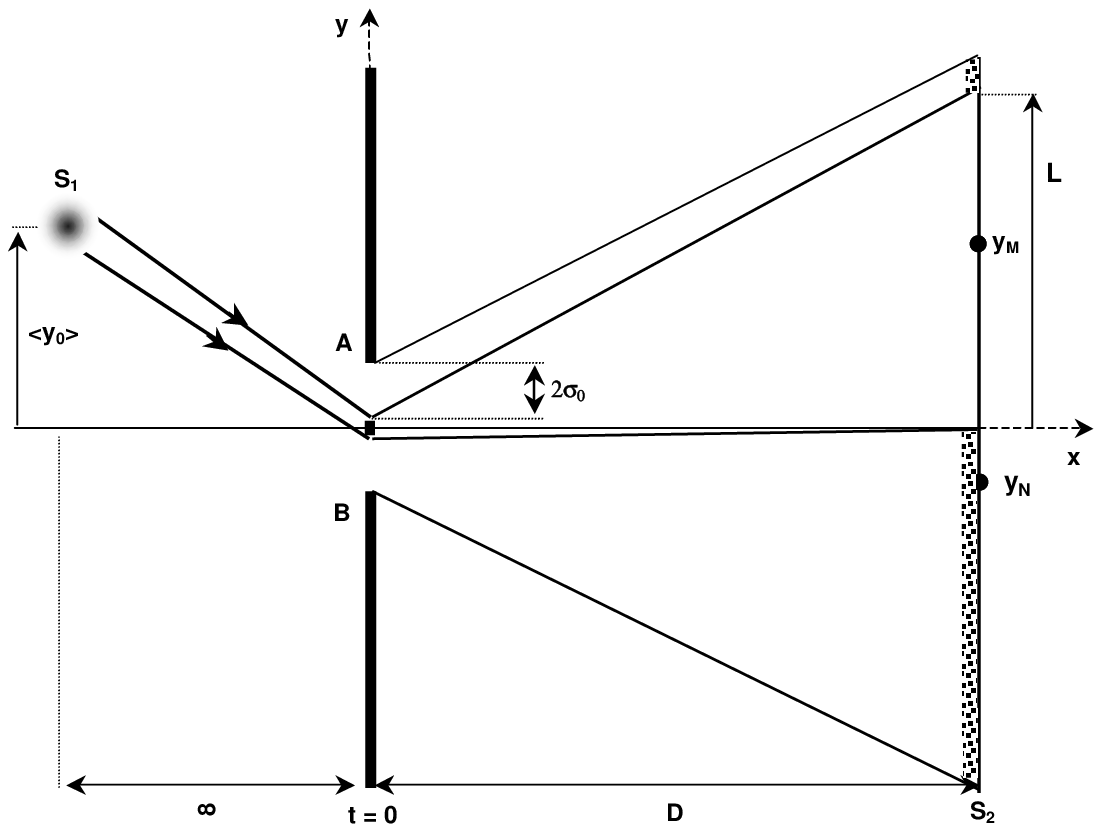}
\caption{Schematic drawing of a two-unentangled particle two-slit
experiment in which the conditions $Y\ll\sigma_0\ll\langle
y_0\rangle$ and $\hbar t/2m\sigma_{0}^{2}\gg 1$ along with
selective detection are considered. The length $L$ shows the empty
interval in the final interference pattern.}
\end{figure}
\end{center}

  However, based on SQM, we have two alternatives:\\
i) The joint probability relation (11) is still valid and there is
only
a reduction in the intensity throughout the screen $S_{2}$, due to the selective detection.\\
ii) SQM is silent about our selective detection.\\
In the first case, there is a disagreement between the predictions
of SQM and BQM and in the second case, BQM has a better predictive
power than SQM, even at the statistical level.

It is worthy to note that, based on our factorizable wave function
(10), one may object that each particle simply follows one of the
single-particle two-slit trajectories and is quite independent of
the other particle and in consequence, both SQM and BQM must
yield the same results.  But, one can see that this objection is
unfounded for our specified conditions in which we have used the
guidance condition along with the selective detection. If we study
the interference pattern without using selective detection, we
must obtain the same results for the two theories. But, using
selective detection, it is clear that not only the two theories
do not have the same statistical predictions, but also BQM
clarifies and illuminates SQM, as Durr et al. [18] said:{\em
``note that by selectively forgetting results we can dramatically
alter the statistics of those that we have not forgotten.  This
is a striking illustration of the way in which Bohmian mechanics
does not merely agree with the quantum formalism, but, eliminating
ambiguities, clarifies, and sharpens it."}.  In our selective
detection, we have forgotten detected single-particle and
two-particle contributions at the one side of the $x$-axis on the
screen $S_{2}$. Elsewhere {\cite{Gols}}, we have also shown that
there is another statistical disagreement between the two
theories for a new two-particle system described by an entangled
wave function, using a different selective detection and without
any deviation of the source from the $x$-axis. Therefore, it
seems that performing such experiments provides observable
differences between the two theories, particularly at the
statistical level.

\section{Conclusion}

In this article, we have suggested two thought experiments to
distinguish between the standard and the Bohmian quantum
mechanics. The suggested experiments consist of a two-slit
interferometer with a special source which emits two identical
non-relativistic particles.  We have shown that, according to the
characteristic of the source, our two-particle system can be
described by two kinds of wave functions: the entangled and the
unentangled wave functions.  For the entangled wave function, we
have obtained some disagreement between SQM and BQM at the
individual level. But, it is shown that, the two theories predict
the same statistical results, as expected. For the unentangled
wave function, the predictions of the two theories could be
different at the individual level too. Again, the results of the
two theories were the same at the ensemble level. However, the
use of selective detection can dramatically alter the
interference pattern, so that not only the statistical results of
BQM do not agree with those of SQM, but BQM can also increase our
predictive power. Therefore, it seems that, our suggested thought
experiments can decide between the standard and the Bohmian
quantum mechanics.

\end{document}